# The determinants of academic career advancement: evidence from Italy[1]

*Giovanni Abramo*[*]
   Laboratory for Studies of Research and Technology Transfer
   Institute for System Analysis and Computer Science (IASI-CNR)
   National Research Council of Italy
   Viale Manzoni 30, 00185 Rome - ITALY
   giovanni.abramo@uniroma2.it

*Ciriaco Andrea D'Angelo*
   Department of Engineering and Management
   University of Rome "Tor Vergata"
   Via del Politecnico 1, 00133 Rome - ITALY
   dangelo@dii.uniroma2.it

*Francesco Rosati*
   Department of Management Engineering
   Technical University of Denmark
   Produktionstorvet  Building 426
   2800 Kgs. Lyngby - Denmark
   frro@dtu.dk

## Abstract

In this work we investigate the determinants of professors' career advancement in Italian universities. From the analyses, it emerges that the fundamental determinant of an academic candidate's success is not scientific merit, but rather the number of years that the candidate has belonged to the same university as the selection committee president. Where applicants have participated in research work with the president, their probability of success also increases significantly. The factors of the years of service and occurrence of joint research for the other commission members also have an effect, however of lesser weight. The specific phenomenon of nepotism, although it exists, seems less important. The scientific quality of the commission members has negligible effect on the expected outcome of the competition, and even more so the geographic location of the university calling for the competition.

# 1. Introduction

Given the role of human capital in the current knowledge-based economy, organizations must attempt to optimize their human resources recruitment and career advancement processes. Such strategies are even more important in the case of higher education systems because of the role universities play in support of national industrial competitiveness, socio-economic development and social mobility.

In competitive higher education systems, universities are in constant competition in seeking out the best researchers and teaching professors from both at home and abroad. However in the higher education systems of several European nations such competitive mechanisms are often weak. In many cases, recruitment and advancement take place by means of relatively rigid procedures, frequently regulated by a central bureaucracy. For example, in Italy appointments to academic positions are not handled through local *ad hoc* search committees or advertisements in international scientific journals; instead all vacancies are submitted to the Ministry of Education, Universities and Research (MIUR) which every so often announces national competitions for all disciplines. Such centrally regulated competitions have come under heavy fire and the Italian word "*concorso*" has soon gained international currency as a term denoting a rigged competition, involving favoritism, nepotism and other unfair selection practices (Gerosa, 2001). A number of instances of court cases (Perotti, 2008; Zagaria, 2007) and injustice have been reported in letters published in such prestigious journals as Lancet, Science and Nature (Garattini, 2001; Aiuti et al., 1994; Biggin, 1994; Amadori et al., 1992; Gaetani and Ferraris 1991; Fabbri, 1986).

Problems of fairness in appointments to academic positions are certainly not limited to Italy: the international literature has dedicated considerable attention to the study of academic recruitment and promotion, largely regarding questions of gender and minority discrimination (Zinovyeva and Bagues, 2012; van den Brink et al., 2010; Cora-Bramble, 2006; Price et al., 2005; Trotman et al., 2002; Stanley et al., 2007). One of the conclusions is that discriminatory phenomena seem to develop above all when evaluations are based on non-transparent criteria (Rees, 2004; Ziegler, 2001; Husu, 2000; Ledwith and Manfredi, 2000; Allen, 1988). In effect, academic recruitment is often reported as an informal process in which a few powerful professors select new ones through cooptation mechanisms (van den Brink et al., 2010; Husu, 2000; Fogelberg et al., 1999; Evans, 1995). Such mechanisms often conceal the phenomenon of favoritism, which has been intensively examined in only a few nations, such as Turkey (Aydogan, 2012), Australia (Martin, 2009), Spain (Zinovyeva and Bagues, 2012) and Italy (Perotti, 2008; Zagaria, 2007). In Italy, there has recently been strong interest in the study of nepotism, which is a particular form of favoritism. While Allesina (2011) and Durante et al. (2011 and 2009) report the unequivocal detection of the phenomenon, Abramo et al. (2014a) are more cautious: while they do not deny the presence of nepotism, they show that the probability of a "child" of a full professor in the same university not meriting his or her position is equal to that of any "non-child". This result is in line with the findings of many sociological studies (Dunn and Holtz-Eakin, 2000; Lentz and Laband, 1989; Simon et al., 1966), which suggest that "children" employed in universities may have in fact received a substantial amount of qualifying career-related knowledge from their parents. Whatever the case for nepotism, it does not deal with the concerns for other forms of favoritism that clearly distort faculty recruitment and career advancement, particularly in countries characterized by scarce intensity of competition among universities. Thus Zinovyeva and Bagues (2012), examining the phenomenon in the Spanish university system, concentrated on the role of connections between the candidates and the evaluators composing the examining boards that decide on academic promotions. They show that the future performance of candidates who were promoted and had a weak connection with the evaluators was better than that of their "non-connected" colleagues. Conversely, successful candidates with a strong link to the evaluators register worse performance both before and after their promotion.

Given the troubling analyses and case evidence, it appears that the efficiency of various recruitment and advancement processes in universities should be subject to ex-post evaluation. After a given lapse of time it is indeed possible to quantify the level of efficiency of the personnel selections through comparative evaluation of the academics' personal achievements in the scientific research sphere. The efficiency of the recruitment and promotion processes can be evaluated by comparing the selected to the rejected candidates in terms of their research performance over a period subsequent to the actual competition.

In a preceding study (Abramo et al., 2014b), the authors investigated the efficiency of the selection process for career advancement in the Italian academic system, referring to the case of all associate professor competitions announced in 2008 (a total of 1,232). This represented a massive intake for a single year, largely for upgrades from assistant to associate professor, with the new staff amounting to 13% of the pre-existing associate professor faculty. The analysis showed that, in the three years subsequent to the competitions, the new associate professors were on average more scientifically productive than their incumbent colleagues. However several critical issues appeared, particularly concerning unsuccessful candidates who outperformed the competition winners in terms of research productivity, as well as a number of competition winners who resulted as totally unproductive. Specifically, it emerged that 29% of the winners had productivity below the median of the performance distribution for their peers in the same field of research, and that 5.5% of the winners had not produced any significant advancement in scientific knowledge. An analysis of the individual competitions showed that almost half of them selected candidates who would go on to achieve below-median productivity in their field of research over the subsequent triennium.

In the present study, given the above critical concerns, the authors investigate the determinants that could have affected the selection procedures other than the scientific merit of the candidates, and attempt to interpret the "non-merit" factors that could potentially have been involved. We investigate the potential phenomenon of favoritism, particularly in terms of the factors of nepotism and social proximity between the candidates and their evaluators. Finally we will examine the correlation between the expected outcomes of competitions and the scientific productivity of the evaluators, as well as the expected outcomes and the territorial localization of the universities that announced the competitions.

We acknowledge that research performance is not the only dimension of quality of a candidate. Evaluations of applicants should also consider dimensions representing the other two institutional missions of universities, meaning teaching and technology transfer. Furthermore, the assessment of research performance by quantity and quality of output alone neglects other attributes of the scientists' activities, for example the ability to manage research teams, to attract funds, their activities in consulting, editorial work, and so on. We would expect some level of correlation between research productivity and these other variables, however caution is recommended in the interpretation of our results, in which we consider only dimensions involving research output.

The next section of the paper summarizes the structure and function of the Italian higher education system, with particular regard to the measures adopted in 2008 for recruitment of associate professors. Sections 3 and 4 present the details of the methodology and dataset used for the analysis, followed by the results. The work concludes with the authors' discussion and recommendations.

## 2. Background

### 2.1 The Italian higher education system

The MIUR recognizes a total of 96 universities as having the authority to issue legally-recognized degrees. Twenty-nine of these are private, small-sized, special-focus universities, of which 13 offer only e-learning. Sixty-seven are public and generally multi-disciplinary universities, scattered throughout the nation. Six of them are *Scuole Superiori* (Schools for Advanced Studies), specifically devoted to highly talented students, with very small faculties and tightly limited enrolment per degree program. In the overall system, 94.9% of faculty are employed in public universities (0.5% in Scuole Superiori). Public universities are largely financed by government through non-competitive allocation. Until 2009 the core government funding (56% of universities' total income) was input oriented, i.e. independent of merit, and distributed to universities in a manner intended to equally satisfy the needs of each and all, in function of their size and disciplines of research. It was only following the first national research evaluation exercise (VTR), conducted between 2004 and 2006, that a minimal share, equivalent to 3.9% of total income, was assigned by the MIUR in function of the assessment of research and teaching.

Despite interventions intended to grant increased autonomy and responsibilities to the universities (Law 168 of 1989[2]), the Italian higher education system is a long-standing, classic example of a public and highly centralized governance structure, with low levels of autonomy at the university level and a very strong role played by the central state.

In keeping with the Humboldtian model, there are no "teaching-only" universities in Italy, as all professors are required to carry out both research and teaching. National legislation includes a provision that each faculty member must provide a minimum of 350 hours per year. At the close of 2012, there were 57,400 faculty members in Italy (full, associate and assistant professors) and a roughly equal number of technical-administrative staff. All new personnel enter the university system through public competitions, and career advancement can only proceed by further public competitions, as indicated in the next section.

---

[2] This law was intended to grant increased autonomy and responsibility to the universities to establish their own organizational frameworks, including charters and regulations. Subsequently, Law 537 (Article 5) of 1993 and Decree 168 of 1996 provided further changes intended to increase university involvement in overall decision-making on use of resources, and to encourage individual institutions to operate on the market and reach their own economic and financial equilibrium.

Salaries are regulated at the centralized level and are calculated according to role (administrative, technical, or professorial), rank within role (for example assistant, associate or full professor) and seniority. None of a professor's salary depends on merit. Moreover, as in all Italian public administration, dismissal of unproductive employees is unheard of.

The entire legislative-administrative context creates an environment and culture that are completely non-competitive, yet flourishing with favoritism and other opportunistic behaviors that are dysfunctional to the social and economic roles of the higher education system. The overall result is a system of universities that are almost completely undifferentiated for quality and prestige, with the exception of the tiny Scuole Superiori and a very small number of private, special-focus universities. The system is thus unable to attract significant talented foreign faculty, or even students. The numbers are negligible: only 1.8% of research staff are foreign nationals. This is a system where every university has some share of top scientists, flanked by another share of absolute non-producers. Over the 2004-2008 period, 6,640 (16.8%) of the 39,512 hard-sciences professors did not publish any scientific articles in the journals indexed by the Web of Science (WoS). Another 3,070 professors (7.8%) did achieve publication, but their work was never cited (Abramo et al., 2013a). This means that 9,710 individuals (24.6%) had no impact on scientific progress measurable by bibliometric databases[3]. An almost equal 23% of professors alone produced 77% of the overall Italian scientific advancement. This 23% of "top" faculty is not concentrated in a limited number of universities, but is instead dispersed more or less uniformly among all Italian universities, along with the unproductive academics, so that no single institution reaches the critical mass of excellence necessary to develop as an elite university and compete at the international level (Abramo et al., 2012a).

## 2.2 Recruitment and career advancement

In Italy, the recruitment and career advancement of professors are regulated by specific law, overseen by the authority of the MIUR. There have been major reforms of the norms over recent years, with the last one being Law 240 of 2010, which introduced a double level of evaluation for selection of associate and full professors. The first level is national, managed directly by the MIUR, and is intended to indicate all those candidates with sufficient qualifications in terms of the scientific activity they have conducted; the second is managed by the individual universities, to select those that are best suited to the specific needs of the university from among those first judged qualified at the national level. Prior to Law 240, the processes of recruitment and career advancement were in the hands of the individual universities, following procedures dictated at the central level. The new two-step national selection procedures are still in a start-up phase.

The last major competition under the old system was held in 2008. In the Italian university system all professors are classified in one and only field (named scientific disciplinary sector - SDS, 370 in all), grouped into disciplines (named university disciplinary areas, UDAs, 14 in all)[4]. In both the new and old system, competitions for recruitment and career advancement occur at SDS level. The 2008 competition procedures required appointment of committees to judge the curricula of the candidates. Each committee was to be composed of five full professors belonging to the SDS for which the position was open. One member, the president, was designated by the university holding the competition and the other four were drawn at random from a short list of other full professors in the SDS concerned. The short list was in turn established by national voting among all the full professors of the SDS. The candidate evaluations carried out by each committee were to be based on:

- examination of the documented qualifications presented by each candidate;
- results of an interview held to better understand the candidate's career profile.

---

[3] Researchers that we define "unproductive" may actually publish in journals not indexed by Web of Science or codify the new knowledge produced in different forms, such as books, patents, etc.

The law required that the selection committees evaluate the candidates' research activity on the basis of specific criteria, including: coherence of the candidate's research history with the scientific field (SDS) for the competition, originality and methodological validity of the scientific production; scientific relevance of the journals or other media of publication and the diffusion of the resulting works within the scientific community; timeliness of the scientific production in relation to the evolution of knowledge in the SDS. The personal documentation to be evaluated was to concern: history of teaching activity; employment service in national and foreign universities and research institutes; organization, direction and coordination of research groups and/or initiatives in teaching and research. As a matter of fact, the discretionary powers of a committee are almost unlimited.

After individual and joint judgments of all the candidates, the committee was required to vote as a whole for selection of two winning candidates[5]. At that point the university that held the competition was free to hire one of the two top finishers for the announced position while the other one remained eligible for hiring over the next five years without further competition, by any other university in the national system.

In order to rationalize the process of the individual competitions over the entire system, the MIUR monitored and gathered the hiring proposals of the various universities and supported the evaluation procedures through information management systems aimed at better guaranteeing transparency. One of the ministry measures was to provide a Web portal[6] with all the basic information on the competition procedures, posts available, number of candidates for each competition, the scheduling of the procedures and final results (winners list, etc.).

The transparency provisions, nomination of a national committee of experts in the field, and the timely issue of regulations for the evaluation procedures were all intended to ensure efficiency in the selection process. In reality, the characteristics of Italian system – such as the generally strong inclination to favoritism, the structured absence of responsibility for poor performance by research units, and the lack of incentive schemes for merit – undermined the credibility of selection procedures for hiring and advancement of university personnel, just as happens for the Italian public administration in general. This is demonstrated by the high and growing number of legal cases brought by losing candidates and by specific studies of systemic problems (Perotti, 2008; Zagaria, 2007), and not least by the unavoidable evidence of that 25% of current Italian faculty are scientifically unproductive.

---

[4] The complete list is accessible on http://attiministeriali.miur.it/UserFiles/115.htm, last accessed on May 26, 2014.
[5] The committees could also indicate a single winner or reject all the applicants, however such events occurred very rarely.
[6] Retrieved from: http://reclutamento.murst.it/, the open Web site managed by the MIUR, titled "Comparative evaluation in the recruitment of University Professors and Researchers (Law 3, 3 July 1998, no. 210)".

## 2.3 Dataset

In 2008, 1,232 competitions for associate professor positions were announced by a total of 74 universities. The competitions concerned 299 SDSs. At the end of all the processes, which lasted an average of over two years[7], the committees had named 2,339 winners, out of a total of 16,500 candidates[8]. The ratio of number of winners of competitions to the size of the existing associate professor faculty averages as 12.8%, varying from a minimum of 8.7% in Earth sciences to a maximum of 21% in Law. In five UDAs the number of applications was higher than the current number of tenured associate professors. For Industrial and information engineering, compared to 1,493 associate professors on faculty, there were 2,010 applications. The competitions generally announced two winners, with the exception of only 39 that announced one. Of the total 1,269 winners, 91.3% (1,159) were academics who were at the time already on staff as assistant professors.

To ensure the representativity of publications as proxy of the research output for bibliometric assessment of research merit of candidates, our analysis focuses only on competitions in the hard sciences, and still more precisely on those SDSs where at least 50% of professors produced at least one publication indexed in the WoS over the period 2004-2008. For this period, there were 654 competitions that met such criteria, in a range of 193 SDSs. The only way to identify all the applicants in these 654 competitions would be to read the minutes of each competition, as were generally published on-line by the individual universities. Given the prohibitive scope of such a task we selected a further subset of 287 competitions (44% of the total 654 in the hard sciences) launched by 12 institutions: the four universities with the highest numbers of competitions from each of the northern, central and southern national areas[9]. For this subset, the winners (550 in all) represent 22% of the total competition candidates (2,590); the rate of selection was more favorable for incumbent assistant professors (532/2,314=23%) than it was for other candidates (18/276=6.5%). Due to the difficulties of authorship disambiguation, our research method is only able to measure the productivity of applicants who are already university faculty members, thus our analysis of career advancement concentrates solely on such candidates. In addition, for the measure of research productivity to be robust, it must be calculated over a sufficiently long period (Abramo et al., 2012b). Because of this, the analysis excludes assistant professors who entered faculty less than three years prior to the date of the competition.

---

[7] At the time of data elaboration, eleven competitions had not been completed.

[8] These figures relate to 1,221 competition procedures that were officially completed (out of 1,232 launched) at the time of preparing the current research paper.

[9] At the time we initiated out study, the minutes of several competitions had been already withdrawn from the various university web sites, however the subset which we ultimately extracted provides a substantial dataset in terms of numerosity of observations and representativity in sectorial and geographic terms. Although the selection was not totally random, there do not appear to be evident problems in generalizing from the analytical results. In fact, given that the commission members are in all cases drawn at random from a national short list and their evaluation procedures are based on a nationally-enforced regulatory structure, it appears reasonable to assume that the phenomena investigated through the subset would be representative of the rest of the national competitions not included in the current observations.

The dataset for the analysis is thus composed of 1,979 assistant professors, 473 of which were competition winners. Table 1 provides the characteristics of our dataset by UDA and its coverage with respect to overall competitions in the 193 SDSs of the hard sciences.

On average there were nine participants per competition, of which eight were Italian-national academics. However the number of candidates shows significant variation (standard deviation 5.6), and 16 competitions involve 20 or more candidates. In the majority of competitions (263 out of 287) the committee designated two winners, with only 24 competitions resulting in a single winner. Table 2 presents the descriptive statistics concerning the candidates.

[Insert Table 1 here]

[Insert Table 2 here]

## 3. The determinants

In this section we present the methods for identifying and measuring the potential determinants of the competition results: scientific merit, nepotism and social proximity between candidates and selection-committee members. The objective is to determine if and to what extent the favoritism-related factors, rather than scientific merit, influence the results.

### 3.1 Measuring scientific merit

Research activity is a production process in which the inputs consist of human resources and other tangible (scientific instruments, materials, etc.) and intangible (accumulated knowledge, social networks, etc.) resources, and where outputs have a complex character of both tangible nature (publications, patents, conference presentations, databases, protocols, etc.) and intangible nature (tacit knowledge, consulting activity, etc.). The new-knowledge production function thus has a multi-input and multi-output character. The principal efficiency indicator of any production system is labor productivity.

The calculation of labor productivity requires a few simplifications and assumptions. It has been shown (Moed, 2005) that in the hard sciences, the prevalent form of codification of research output is publication in scientific journals. As a proxy of total output, in this work we consider only the specific publications (articles, article reviews, and proceeding papers) indexed in the Thomson Reuters WoS.

When measuring labor productivity, if there are differences in the production factors available to each scientist then one should normalize by them. Unfortunately relevant data are not available at the individual level in Italy. The first assumption then is that resources available to professors within the same field are the same. The second assumption is that the hours devoted to research are more or less the same for all professors. Given the characteristics of the Italian academic system as depicted in section 2.1, the above assumptions appear acceptable.

Because of the differences in the publication intensity across fields, a prerequisite of any distortion-free performance assessment is the classification of each researcher in one and only one field (Abramo et al., 2013b).

Most bibliometricians define productivity as the number of publications in the period of observation. Because publications have different values (impact), we prefer to adopt a more meaningful definition of productivity, i.e. the value of output per unit value of labor, all other production factors being equal. The latter recognizes that the publications embedding new knowledge have different value or impact on scientific advancement, which bibliometricians approximate with citations or journal impact factors. Provided that there is an adequate citation window (at least two years) the use of citations is always preferable (Abramo et al., 2011). Because citation behavior varies by field, we standardize the citations for each publication with respect to the average of the distribution of citations for all the Italian cited publications of the same year and the same WoS subject category[10]. Furthermore, research projects frequently involve a team of professors, which is registered in the co-authorship of publications. In this case we account for the fractional contributions of scientists to outputs, which is at times further signaled by the position of the authors in the byline.

At the individual level, professors of the same academic rank in this specific case, we can measure the average yearly productivity, named Fractional Scientific Strength (*FSS*), in the following way[11]:

$$FSS = \frac{1}{t} \sum_{i=1}^{N} \frac{c_i}{\bar{c}} f_i$$

[1]

Where:
t = number of years of work of the researcher in the period of observation;
N = number of publications of the researcher in the period of observation;
$c_i$ = citations received by publication *i*;
$\bar{c}$ = average of the distribution of citations received for all cited publications of the same year and subject category of publication *i*;
$f_i$ = fractional contribution of the researcher to publication *i*.

Fractional contribution equals the inverse of the number of authors in those fields where the practice is to place the authors in simple alphabetical order but assumes different weights in other cases. For the life sciences, widespread practice in Italy is for the authors to indicate the various contributions to the published research by the order of the names in the byline. For the life science SDSs, we give different weights to each co-author according to their order in the byline and the character of the co-authorship (intra-mural or extra-mural) (see Abramo et al., 2013c). If first and last authors belong to the same university, 40% of citations are attributed to each of them; the remaining 20% are divided among all other authors. If the first two and last two authors belong to different universities, 30% of citations are attributed to first and last authors; 15% of citations are attributed to second and last author but one; the remaining 10% are divided among all others[12].

---

[10] Abramo et al. (2012c) demonstrated that the average of the distribution of citations received for all cited publications of the same year and subject category is the best-performing scaling factor.
[11] A more extensive theoretical dissertation on how to operationalize the measurement of productivity can be found in Abramo and D'Angelo (2014a).

Data on faculty of each university and their SDS classification are extracted from the database on Italian university personnel, maintained by the MIUR. The bibliometric dataset used to measure *FSS* is extracted from the Italian Observatory of Public Research (ORP), a database developed and maintained by the authors and derived under license from the Thomson Reuters WoS. Beginning from the raw data of the WoS, and applying a complex algorithm for reconciliation of the author's affiliation and disambiguation of the true identity of the authors, each publication (article, article review and conference proceeding) is attributed to the university scientist or scientists that produced it (D'Angelo et al., 2011). Thanks to this algorithm we can produce rankings of research productivity at the individual level, on a national scale. Based on the value of *FSS* we obtain, for each SDS, a ranking list expressed on a percentile scale of 0-100 (worst to best) for comparison with the performance of all Italian colleagues of the same academic rank and SDS

## 3.2. Nepotism

Non-competitive higher education systems are exposed to greater risks of favoritism in recruitment and career advancement. In countries such as Italy, where political and economic favoritism are widespread, favoritism phenomena can be more accentuated (Zagaria, 2007; Perotti, 2009). The authors have previously explored the aspect of nepotism in the Italian university system (Abramo et. al., 2014a), as noted in the introduction. The previous research revealed that nepotism does take place, as seen in cases of children who advanced in their careers despite scientific performance in the bottom 20% of their SDSs. Such events were observed in 7.8% of the total of children who advanced in rank.

In this work we will again identify the assistant professor candidates in our dataset who presumably have a parent full professor in the same university and in the same year as for their competition.

The starting point for identifying family links within the same university is the identification of professors with the same last name (Allesina, 2011; Durante et al., 2009; Angelucci et al. 2010; Güell et al., 2007). Pairs are then identified among the homonymous professors. For convenience we label the pairs as "parent-child," even if they could be grandparent-grandchild, uncle/aunt-nephew/niece, brother-sister, cousins, or unrelated. This procedure inevitably excludes identification of most family relations headed by the "mother". To make the identification of professors potentially subject to nepotism more reliable we impose the following conditions. The field of observation for "children" concerns assistant professors who participated in a competition announced in the year 2008; the "parents" are the full professors of the same university as the child, in role in the year 2008. Further, the family name must not be included on the list of the 500 most common surnames in Italy[13], nor among the 20 most common in the region where the university is based[14].

---

[12] The weighting values were assigned following advice from senior Italian professors in the life sciences. The values could be changed to suit different practices in other national contexts.

[13] Retrieved from the "International Laboratory of Onomastics".
http://onomalab.uniroma2.it/contents/allegati/3000_cognomi_italia_2000.pdf. Last accessed on May 26, 2014.

Beginning from the starting dataset of 1979 assistant professors, we identify 81 children (4.1% of the total candidates), of which 24 were competition winners. The rate of selection was more favorable for children (24/81 = 29.6%) than it was for non-children (449/1,898 = 23.7%).

### 3.3 Social proximity

The influence of applicant to evaluator social proximity on the outcome of competitions was previously examined by Zinovyeva and Bagues (2012). Their research concentrated on the connections between candidates and examining-board evaluators for academic promotion in Spain. They consider various types of links, distinguishing strong ties (scientific collaboration, advisory roles, belonging to the same university) from weak ties and indirect ties,[15] and show that the future performance of candidates who were promoted and had a weak connection with the evaluators was better than that of their "non-connected" colleagues. Conversely, successful candidates with a strong link to the evaluators register worse performance both before and after their promotion.

In the current study we investigate the influence of social proximity on career advancement of Italian academics, distinguishing two types of candidate-evaluator links:

- Workplace links: the candidate is from the same university as the committee president or as one of the other members. The intensity of the link is expressed by the number of the years that the applicant has spent in the same university and same SDS as the committee president or as the other evaluators, in the period 2001-2010;

- Professional links: the candidate has co-authored publications with the committee president or members. For committee presidents, the link intensity is given by the percentage of the president's publications coauthored with the candidate out of the total of the president's publications in the period 2001-2010. For the committee, intensity is given by the number of committee members who co-authored 2001-2010 publications with the candidate (value from 0 to 4).

The applicant's career history is a fundamental consideration because it reveals his or her experience and qualifications. It is also important because career events include development of a network of contacts and colleagues that permit future benefits of various kinds, from professional development opportunities to favoritism in access to resources and career progression. We thus examine the careers of both the applicants and evaluators to determine the number of years that each applicant spent in the same university and SDS as his or her evaluators over the period 2001-2010. We include the "post-competition" years 2009-2010 because the completion of the announced competitions in almost all cases extended through these additional two years. We also distinguish the career years shared with the committee president from those shared with the other evaluators, based on the hypothesis that the president, chosen by the university that called for the competition, will have greater weight in the final committee decision than the other evaluators. Instances of shared professional work can be objectively measured by the proxy of publications in co-authorship. Cooperation in common projects, where scientists have common research objectives and interests, can in fact lead to very strong social ties characterized by reciprocal opinions of respect and trust. To analyze the influence of such candidate-evaluator research collaborations on competition outcomes we selected the publications for the 2001-2010 period. Although the competitions were announced in 2008, we again included publications from 2009-2010, in this case because substantial time can elapse before a publication actually arrives in print. An article published as late as 2010 is very likely the result of shared work in previous years, including 2008 and prior. This measure also again takes account of the fact that the first steps in the candidate evaluations were carried out a year after the competition announcement, and thus could be influenced by post-2008 collaborations.

---

[14] Retrieved from www.cognomix.it. Last accessed on May 26, 2014.

[15] According to Zinovyeva and Bagues (2012), a weak ties exists when at least one of the following conditions occurs: the evaluator was a member of the candidate's thesis committee; the evaluator has invited the candidate to sit on the thesis committee of one of his or her students (or vice versa); the evaluator and the candidate sat on the same thesis committee. An indirect tie exists when at least one of the following conditions occurs: the evaluator and the candidate have either a common advisor or a common thesis committee member or a common co-author.

## 4. Statistical Analysis

For the current study we formulated a statistical model that links the competition outcome to the determinants described in Section 3.

The dependent variable, the competition outcome, is a Boolean type variable with value of 1 in the case that the applicant wins, or 0 otherwise. The six independent variables are: the parental link between applicant and full professors in the same university; the career years that an applicant has spent in the same university and same SDS as the committee president; the career years that an applicant has spent in the same university and the same SDS as other committee members; the percentage of the president's publications coauthored with the candidate; the number of other committee members with which the applicant has co-authored publications; and, as proxy of scientific merit, the applicant's scientific productivity FSS for the five years 2004-2008 (with citations observed on 31/12/2011).

As the basis of the statistical model we choose the logistic regression function (rendered linear through the *logit* function), which is particularly suited for modeling dichotomous dependent variables. A strictly linear model would not be sufficiently representative of the distribution of values.

Formally, the statistical model is described:

$$logit(p) = \beta_0 + \beta_1 FSS + \beta_2 NE + \beta_3 CP + \beta_4 CM + \beta_5 PP + \beta_6 MP$$

[2]

Where:

$$logit(p) = log \frac{p(E)}{1-p(E)}$$

$E$ = competition outcome, with a value of 1 if the applicant wins the competition, otherwise 0;

$p(E)$ = probability of event E;

$\beta$ = generic regression coefficient;

$FSS$ = applicant's research productivity over the period 2004-2008, expressed on a 0-100 percentile scale;

$NE$ = value of 1 if the applicant and a full professor in the same university have the same family name, as of 31/12/2008; otherwise 0;

$CP$ = applicant's career years in the same university and same SDS as the committee president over the period 2001-2010;

$CM$ = applicant's career years in the same university and the same SDS as the other evaluation committee members over the period 2001-2010.

$PP$ = percentage of committee president's publications in co-authorship with the candidate over the period 2001-2010.

$MP$ = number of other committee members with which the applicant has co-authored publications over the period 2001-2010.

Prior to applying the statistical model we present the descriptive statistics for the variables in Table 3. For each variable we show the average, standard deviation (SD) and the maximum value occurring for the competition winners, non-winners and total applicants in the dataset.

[Insert Table 3 here]

As would be expected, the winners' scientific performance is on average higher than that of non-winners (68.29 for winners versus 61.73 for non-winners). However the statistics also permit several other interesting observations. For all applicants, the average number of years in the same university as the committee president is 2.11; for winners this figure rises to 4.27 and for non-winners it drops to 1.43. For the set of all applicants, the average number of years spent in the same university as the other evaluators is 1.27, compared to 1.23 for the winners and 1.28 for non-winners. Concerning publications, on average the full set of participants contribute to 2.37% of the president's scientific production; winners contribute to 7.01% and non-winners to 0.91%. Among winners, 5% have the same family name of a full professor in the same university; among non-winners, 4%.

## 4.1 Correlation Analysis

Table 4 presents the results from the test for links between the regressors. In this case, the Pearson correlation analysis indicates the absence of correlation between the independent variables figuring in the model. The highest correlation is between PP and CP, at 0.365. This is line with what we expect, since scientists in the same university and SDS would tend to cooperate in shared research work.

Thus from the Pearson correlation analysis it emerges that the hypothesis of independence between the variables can be considered valid.

[Insert Table 4 here]

## 4.2 The Logistic Regression Model

Table 5 presents the logistic regression results predicting the competition outcomes.

[Insert Table 5 here]

The odds ratio for the competition outcomes (i.e. probability of winning the competition relative to the probability of not winning) is formalized:

$$\frac{p(E)}{1-p(E)} = \exp(-2.696 + 0.013 * FSS + 0.468 * NE + 0.178 * CP + 0.036 * CM + 0.030 * PP + 0.618 * MP)$$

[3]

The value $e^b$, calculated for each potential explanatory variable, represent OR in Table 1. Where OR equals 1 the associated explanatory variable would have no effect on the dependent variable of "competition outcome".

The average value of VIF (Variance inflation factor), calculated at 1.11, confirms the absence of multicollinearity in the model.

The values calculated for standardized $b$ (last column, Table 5) permit comparison of the effects of the variables measured in different metrics. The data indicate that the factor having the greatest influence on the competition outcomes ($b\text{Std}_{CP}$=0.655) seems to be the number of the applicant's years in the same university and same SDS as the committee president. In particular, every unit increase in the number of career years shared with the president increases the odds of success by a factor of 1.195. Co-authorship of publications with the committee president (PP) also has remarkable bearing on the competition results ($b\text{Std}_{PP}$=0.313), with every percent increase in PP increasing the odds of success by a factor of 1.030. As well, the applicant's scientific productivity (FSS) has notable weight ($b\text{Std}_{FSS} = 0.332$), with every unit increase in the FSS increasing the odds of success by a factor of 1.013. Shared research work with other committee members has a lesser bearing ($b\text{Std}_{MP} = 0.192$) on competition outcome, however still to a significant level. Similarly, the applicant's career years with the other evaluators ($b\text{Std}_{CM} = 0.121$) also has lesser bearing, and to a lower significant level. Concerning the parental link between applicant and full professors in the same university (NE), the small number of "children" makes this finding non-significant.

Returning to the formula model, we can now calculate $p$, the probability of succeeding in career progression competitions based on the specific observed values of the independent variables:

$$p(E=1) = \frac{\exp(b_0 + b_1 FSS + b_2 NE + b_3 CP + b_4 CM + b_5 PP + b_6 MP)}{1 + \exp(b_0 + b_1 FSS + b_2 NE + b_3 CP + b_4 CM + b_5 PP + b_6 MP)}$$

[4]

Table 6 provides the results from simulations in which we estimate the probability of a candidate's success based on different values of FSS and CP, assuming mean values for all other variables. From the model we see that a candidate with the maximum possible FSS (percentile 100) but without any shared history in the committee president's university would have a 0.231 probability of success. However a candidate with a value of FSS equal to 50 but with five years in the same university and SDS as the president would have better chances of succeeding (equal to 0.276).

[Insert Table 6 here]

## 5. Impact of evaluators' scientific productivity, the geographic area of competition, and international reputation in the discipline

In this section we examine the impact on competition outcomes of: i) the committee members' scientific quality; ii) the geographic area of the selection competitions; and iii) the international reputation of Italy in the discipline.

### 5.1 Impact of the evaluators' scientific productivity on competition outcomes

In analyzing the impact of the evaluators' scientific productivity, we consider the competition outcome: as expected, if the winners' FSS over the period 2004-2008 results as not less than the median of FSS distributions for both the competition applicants and for all assistant professors in the same SDS; otherwise not as expected. The committees are considered: adequate, if at least three members out of five show a 2004-2008 FSS not less than the median of the distribution of all full professors in the SDS: otherwise inadequate.

To ensure a robust measure of the candidates' research productivity the analysis excludes those competitions that fail to show a winner or any non-winning participant with at least three years on staff in the 2004-2008 period (46 out of 287). Similarly, regarding the productivity of the evaluators, we exclude the competitions where one or more committee members had held their faculty role for less than three years over the 2004-2008 period (2 out of 287) (see Abramo et al., 2012b). Given the exclusions, we reduce the number of competitions observed to 239. Of these there are 106 competitions where the outcomes are as expected, of which 74 had adequate committees, and 133 with non-expected outcomes, of which 77 had adequate committees (Table 7). The test for association shows a Pearson chi-square result of 3.60, with a non-significant p-value of 0.058, and a likelihood-ratio chi-square of 3.63, with non-significant p-value of 0.057. The odds ratio (OR) results as 1.68, demonstrating a modest positive association between competition's with expected outcomes and adequacy of the scientific committee.

[Insert Table 7 here]

### 5.2 Influence of the geographic area

A frequently encountered opinion is that favoritism in academic institutions is more concentrated in southern than northern Italy (Allesina, 2011; Durante et al., 2011). We thus localize the universities (north, central, south) that announced the competitions and conduct an analysis to detect the potential influence of geographic area on their outcome.

The north-south comparison shows that there are 84 competitions with expected outcomes (applying the same criteria for winners' scientific productivity as in the preceding section), of which 58 were held in the north, and 104 competitions with non-expected outcomes of which 68 were held in the north (Table 8). The Pearson chi-square results as 0.28, with a non-significant p-value at 0.595, and the likelihood-ratio chi-square is 0.28 with a non-significant p-value at 0.595. The odds ratio (OR) calculated for the association is 1.18, indicating a weakly positive association between the outcome of an "expected" competition result and the fact that the competition is held in northern rather than southern university.

[Insert Table 8 here]

We also conduct the same analysis for the north-central and central-south combinations of competition location. For the north to center comparison, the Pearson chi-square result is 0.429, with a non-significant p-value of 0.513. The likelihood-ratio chi-square result is 0.430 with a non-significant p-value of 0.512. The odds ratio (OR) results as 1.24, indicating a weak positive association between the outcome of an "expected" competition result and the fact that the competition is held in a northern rather than central-Italian university. In the case of the central to south comparison, both the Pearson chi-square and likelihood-ratio chi-squared give results of 0.017, with non-significant p-value at 0.896. The odds ratio is 0.95, showing a very weak association between competition outcomes "as expected" and the fact that the competition is held by a central-Italian rather than southern university.

### 5.3 Impact of the international reputation of Italy in the discipline

One would expect that committees in disciplines where Italy has an international long-standing reputation would be more reluctant than others to sacrifice research excellence to other dimensions of quality in the selection criteria. We tested this hypothesis, starting from the results of a recent investigation by Abramo and D'Angelo (2014b), which showed that Italy's relative scientific strength is high in physics, chemistry, and medicine, and low in economics and statistics, civil engineering, and mathematics. We then analyzed the distributions of competition outcomes by UDA (Table 9).

[Insert Table 9 here]

The analysis shows that mathematics and earth sciences are the only UDAs where the majority of outcomes are as expected. We conducted association tests between the competition outcomes in the UDA of mathematics and, respectively, physics, chemistry and medicine. Between mathematics and physics, the test for association shows Pearson chi-square = 4.29 with a p-value = 0.037, likelihood-ratio chi-square = 4.33 with a p-value = 0.038. The odds ratio (OR) = 3.82, demonstrates a more than moderate positive association between competitions with expected outcomes and the fact that the competitions occurred in mathematics rather than in physics. The same holds true for the association tests between mathematics and chemistry (Pearson chi-square = 5.50 with a p-value = 0.019, likelihood-ratio chi-square = 5.61 with a p-value = 0.018, OR = 4.25); and between mathematics and medicine (Pearson chi-square = 6.17, with p-value = 0.013, and a likelihood-ratio chi-square = 6.31, with p-value = 0.012, OR = 3.64. A relationship between international scientific strength in a discipline and attention to the scientific merit of candidates can then be excluded.

## 6. Discussion and conclusions

The presence of competition is known to serve as a major stimulus for continuous improvement and the pursuit of excellence. Lacking such stimuli, non-competitive education systems appear more at risk of non-efficient recruitment and career advancement processes, with negative outcomes for quality of education and research. Such phenomena appear even more probable in countries with a general environment characterized by favoritism.

In a preceding work (Abramo et al., 2014b), we evaluated the efficiency of the selection process for career advancement of university professors, referring to the case of competitions for associate professor announced in the Italian academic system for the year 2008. The analyses showed that in the three years following the competitions, the new associate professors were on average more productive than their incumbent colleagues. However several critical issues appeared, particularly concerning unsuccessful candidates who outperformed the competition winners in terms of productivity over the subsequent triennium, as well as a number of competition winners who resulted as totally unproductive. An analysis of the individual competitions showed that almost half of the selected candidates would go on to achieve below-median productivity in their field of reference over the subsequent period.

In the present study, we have investigated the determinants of the results emerging from the previous study. Our intention was to provide an interpretation of the potential factors that could have contributed to the outcomes of the 2008 round of competitions. In particular, we investigated the extent to which: i) favoritism and nepotism could have conditioned the competition outcomes; ii) if the scientific quality of the selection committee members could have influenced outcomes; iii) if there are significant differences in outcomes between broad geographic areas.

From the analyses, it emerges that the fundamental determinant of a candidate's success is not his or her scientific merit, but rather the number of their years of service in the same university as the committee president. Where the candidate has cooperated in joint research work with the president the probability of success also increases significantly. The factors of the years of service and occurrence of joint research with the other committee members have lesser weight. The phenomenon of potential nepotism, although it occurs, seems to have a lower impact. These outcomes explain also why universities are closed shops; the concentration index of incumbent assistant professors who win the competitions is 1.1 against 0.3 of outside candidates.

The results of our regression model confirm what Zinovyeva and Bagues (2012) have demonstrated for the Spanish case, showing once again that links between colleagues have a strong weight in the final judgment of the selection committees. In fact we observe that when the candidate has such links with the committee president the influence on the competition outcome seems even greater than the weight of the applicant's scientific productivity.

A further step in the analysis detected that there is a modest positive association between an "expected" outcome to the competition, where the winner truly has scientific merit, and the scientific value of the committee members. A final test for concentration of potential favoritism in particular national areas showed that there is a weak positive association between "expected" outcomes to competitions and the fact that these are held in northern universities rather than southern or central-Italian universities.

A scientific standpoint cannot do without warning the decision maker about the limits of the analysis and suggesting the usual caution in the interpretation of results. First of all, all limits of inferential analysis apply here: our results are based on a subset of 243 observations out of a population of 654 competitions in the hard sciences. To establish the research strength of candidates we have measured their productivity in the period 2004-2008. Research performance before 2004 may well have been taken into account by the selection committees. Furthermore, it should be recognized that scientific merit is not the only dimension of the quality of a candidate. Other dimensions may also have been considered by the selection committees, such as teaching skills, technology transfer achievements, demonstrated abilities to attract funds, coordinate research teams or carry out editorial activities. Evaluators might in principle be more informed about the true quality of connected candidates, and they might efficiently use their private information in their evaluations. There is some likelihood that the candidate with the best fit to a given department, is already employed by it.

The results of our analysis may be open then to a twofold interpretation: favoritism or "smart" recruitment. While on the one side findings on the future productivity of candidates (Abramo et al., 2014b) seem to exclude efficiency in the use of inside information by the committee members, on the other side the fact that in disciplines with strong scientific performance competitions outcomes as expected are fewer than in weak disciplines seems to suggest smart recruitment. Further research is needed to find out which is the correct interpretation of our findings, which may also subtend both favoritism in some competitions and smart hiring in others,

## References

Abramo, G., D'Angelo, C.A., Rosati, F. (2014a). Relatives in the same university faculty: nepotism or merit? *Scientometrics*. DOI: 10.1007/s11192-014-1273-z.

Abramo, G., D'Angelo, C.A., Rosati F., (2014b). Career advancement and scientific performance in universities. *Scientometrics*. 98(2), 891-907

Abramo, G., D'Angelo, C.A. (2014a). How do you define and measure research productivity? *Scientometrics*. DOI: 10.1007/s11192-014-1269-8.

Abramo, G., D'Angelo, C.A. (2014b). Assessing national strengths and weaknesses in research fields. *Journal of Informetrics,* 8(3), 766–775.

Abramo, G., Cicero, T., D'Angelo, C.A., (2013a). The impact of non-productive and top scientists on overall university research performance. *Journal of Informetrics*, 7(1), 166-175.

Abramo, G., Cicero, T., D'Angelo, C.A. (2013b). Individual research performance: a proposal for comparing apples to oranges. *Journal of Informetrics*, 7(2), 528-539.

Abramo, G., D'Angelo, C.A., Rosati, F. (2013c). The importance of accounting for the number of co-authors and their order when assessing research performance at the individual level in the life sciences. *Journal of Informetrics*, 7(1), 198-208.

Abramo, G., Cicero, T., D'Angelo, C.A., (2012a). The dispersion of research performance within and between universities as a potential indicator of the competitive intensity in higher education systems. *Journal of Informetrics*, 6(2), 155-168.

Abramo, G., D'Angelo, C.A., Cicero, T. (2012b). What is the appropriate length of the publication period over which to assess research performance? *Scientometrics*, 93(3), 1005-1017.

Abramo, G., Cicero, T., D'Angelo, C.A. (2012c). Revisiting the scaling of citations for research assessment. *Journal of Informetrics*, 6(4), 470–479.

Abramo, G., Cicero, T., D'Angelo, C.A. (2011). Assessing the varying level of impact measurement accuracy as a function of the citation window length. *Journal of Informetrics*, 5(4), 659-667.

Aiuti, F, Bruni, R., Leopardi, R. (1994). Impediments of Italian science. *Nature*, 367(6464), 590-590.

Allen, N. (1988). Aspects of promotion procedures in Australian universities. *Higher Education*, 17(3), 267-280.

Allesina, S. (2011). Measuring nepotism through shared last names: The case of Italian academia. *PlosONE*, 6(8), e21160.

Amadori, S., Bernasconi, C., Boccadoro, M., Glustolisi, R., Gobbi, M. (1992). Academic promotion in Italy. *Nature*, 355(6361), 581-581.

Angelucci, M., De Giorgi, G., Rangel, M.A., Rasul, I. (2010). Family networks and school enrolment: Evidence from a randomized social experiment. *Journal of Public Economics*, 94(3), 197-221.

Aydogan, I. (2012). The existence of favoritism in organizations. *African Journal of Business Management*, 6(12), 4577-4586.

Biggin S. (1994). Corruption scandal reaches academe. *Science*, 266(5187), 965-965.

Cora-Bramble, D. (2006). Minority faculty recruitment, retention and advancement: Applications of a resilience-based theoretical framework. *Journal of Health Care* *for the Poor and Underserved*, 17(2), 251-255.

D'Angelo, C.A., Giuffrida, C., Abramo, G. (2011). A heuristic approach to author name disambiguation in large-scale bibliometric databases. *Journal of the American Society for Information Science and Technology*, 62(2), 257–269.

Dunn, T., Holtz Eakin, D. (2000). Financial capital, human capital, and the transition to self-employment: evidence from intergenerational links. *NBER Working Paper* No. 5622.

Durante, R., Labartino, G., Perotti, R. (2011). Academic dynasties: decentralization and familism in the Italian academia. *NBER Working Paper* No. 17572.

Durante, R., Labartino, G., Perotti, R., Tabellini, G. (2009). Academic Dynasties. *Bocconi University Working Paper*.

Evans, C. (1995). Choosing people: recruitment and selection as leverage on subjects and disciplines. *Studies in Higher Education*, 20(3), 253-265.

Fabbri, L.M. (1987). Rank injustice and academic promotion. *Lancet*, 2(8563), 860-860.

Fogelberg, P., Hearn, J., Husu, L., Mankkinnen, T. (1999). Hard Work in the Academy: Research and Interventions on Gender Inequalities in Higher Education. *Helsinki University Press*, POB 4 (Vuorikatu 3), FIN-00014 University of Helsinki. ISBN: 978-9-5157-0456-6.

Gaetani, G.F., Ferraris, A.M. (1991). Academic promotion in Italy. *Nature*, 353 (6339), 10-10.

Garattini, S. (2001). Competition for academic promotion in Italy – Reply. *Lancet*, 357(9263), 1208-1208.

Gerosa, M. (2001). Competition for academic promotion in Italy. *Lancet*, 357(9263), 1208-1208.

Güell, M., Rodríguez Mora, J., Telmer, C. (2007). Intergenerational mobility and the informative content of surnames. *CEPR Discussion Paper*, 6316.

Husu, L. (2000). Gender discrimination in the promised land of gender equality. *Higher Education in Europe*, 25(2), 221-228.

Ledwith, S., Manfredi, S. (2000). Balancing Gender in Higher Education A Study of the Experience of Senior Women in a `New' UK University. *European Journal of Women's Studies*, 7(1), 7-33.

Lentz, B.F., Laband, D.N. (1989). Why so many winners of doctors become doctors: nepotism vs. human capital transfers, *The Journal of Human Resources*, 24(3), 396-413.

Martin, B. (2009). Academic patronage. *International Journal for Educational Integrity*, 5(1), 3-19.

OECD (2013). OECD Skills Outlook 2013: First Results from the Survey of Adult Skills, OECD Publishing. ISBN 978-92-64-20425-6. DOI: 10.1787/9789264204256-en

Moed, H.F. (2005). *Citation Analysis in Research Evaluation*. Springer, ISBN: 978-1-4020-3713-9.

Perotti, R. (2008). *L'università truccata*. Einaudi, Torino, Italy. ISBN: 978-8-8061-9360-7.

Price, E.G., Gozu, A., Kern, D.E., Powe, N.R., Wand, G.S., Golden, S., Cooper, L.A. (2005). The role of cultural diversity climate in recruitment, promotion, and retention of faculty in academic medicine. *Journal of general internal medicine*, 20(7), 565-571.

Rees, T. (2004). Measuring excellence in scientific research: the UK Research

Assessment Exercise. In *Gender and excellence in the making* (pp. 117-123). European Commission, Brussels, Belgium. ISBN: 92-894-7479-3.

Schwab, K. (2012). *The Global Competitiveness Report 2012–2013*. Report of the World Economic Forum. ISBN 92-95044-35-5

Simon, R. J., Clark, S.M. , Tifft, L.L. (1966). Of Nepotism, Marriage, and the Pursuit of an Academic Career. *Sociology of Education*, 39(4), 344-358.

Stanley, J.M., Capers, C.F., Berlin, L.E. (2007). Changing the face of nursing faculty: minority faculty recruitment and retention. *Journal of Professional Nursing*, 23(5), 253-261.

Trotman, C.A., Bennett, E., Scheffler, N., Tulloch, J.C. (2002). Faculty recruitment, retention, and success in dental academia. *American journal of orthodontics and dentofacial orthopedics*, 122(1), 2-8.

Van den Brink, M., Benschop, Y., Jansen, W. (2010). Transparency in academic recruitment: a problematic tool for gender equality? *Organization Studies*, 31(11), 1459-1483.

Zagaria, C. (2007). *Processo all'università. Cronache dagli atenei italiani tra inefficienze e malcostume*. Dedalo, Bari, Italy. ISBN: 978-8-8220-5365-7.

Ziegler, B. (2001). Some remarks on gender equality in higher education in Switzerland. *International journal of sociology and social policy*, 21(1/2), 44-49.

Zinovyeva, N., Bagues, M. (2012). The role of connections in academic promotions. *Business Economics Working Papers* from Universidad Carlos III, Instituto sobre Desarrollo Empresarial "Carmen Vidal Ballester". Available at SSRN 2136888, http://papers.ssrn.com/sol3/papers.cfm?abstract_id=2136888, last accessed on May 26, 2014.

***Table 1: Population subset selected for analysis (in parentheses the percentage with respect to the overall reference population)***

| UDA | Competitions | SDSs concerned | Universities launching competitions | Winners | Academic winners with seniority $\geq 3$ years |
|---|---|---|---|---|---|
| Mathematics and computer science | 26 (46%) | 7 (78%) | 10 (40%) | 50 (46%) | 45 (47%) |
| Physics | 19 (42%) | 5 (63%) | 8 (33%) | 37 (43%) | 30 (41%) |
| Chemistry | 25 (46%) | 8 (67%) | 9 (38%) | 47 (46%) | 44 (48%) |
| Earth sciences | 6 (30%) | 4 (33%) | 4 (33%) | 10 (27%) | 5 (17%) |
| Biology | 25 (34%) | 14 (74%) | 10 (30%) | 49 (34%) | 39 (31%) |
| Medicine | 62 (41%) | 32 (68%) | 9 (27%) | 116 (40%) | 87 (40%) |
| Agricultural and veterinary sciences | 15 (31%) | 11 (39%) | 5 (26%) | 27 (29%) | 26 (30%) |
| Civil engineering and architecture | 11 (42%) | 6 (86%) | 8 (38%) | 22 (43%) | 22 (46%) |
| Industrial and information engineering | 86 (60%) | 31 (74%) | 12 (39%) | 170 (62%) | 155 (62%) |
| Pedagogy and psychology | 5 (24%) | 3 (60%) | 3 (23%) | 8 (20%) | 7 (21%) |
| Economics and statistics | 7 (39%) | 3 (75%) | 4 (36%) | 14 (39%) | 13 (41%) |
| Total | 287 (44%) | 124 (64%) | 12 (21%) | 550 (43%) | 473 (44%) |

***Table 2: Descriptive statistics for the candidates involved in the dataset of competitions***

| | Winners | Non winners | Total | Candidates per competition | | | |
| --- | --- | --- | --- | --- | --- | --- | --- |
| | | | | Average | Median | Std Dev | Max |
| Total candidates | 550 | 2,040 | 2,590 | 9 | 8 | 5.6 | 29 |
| Academics | 532 | 1,782 | 2,314 | 8 | 7 | 5.4 | 28 |
| Others | 18 | 258 | 276 | 1 | 1 | 1.2 | 6 |
| Academics with seniority ≥ 3 years | 473 | 1,506 | 1,979 | 7 | 6 | 4.6 | 26 |
| Academics with seniority < 3 years | 59 | 276 | 335 | 1 | 1 | 1.3 | 6 |

**Table 3: Descriptive statistics for logistic regression variables**

| Var. | Winners | | | Non winners | | | Total | | |
|------|-----|-----|-----|-----|-----|-----|-----|-----|-----|
| | Avg | SD | Max | Avg | SD | Max | Avg | SD | Max |
| FSS | 68.29 | 23.37 | 100 | 61.73 | 25.02 | 100 | 63.30 | 24.79 | 100 |
| NE | 0.05 | 0.22 | 1 | 0.04 | 0.19 | 1 | 0.04 | 0.20 | 1 |
| CP | 4.27 | 4.37 | 10 | 1.43 | 3.15 | 10 | 2.11 | 3.68 | 10 |
| CE | 1.23 | 3.48 | 20 | 1.28 | 3.37 | 21 | 1.27 | 3.39 | 21 |
| PP | 7.01 | 17.01 | 91.18 | 0.91 | 6.23 | 97.36 | 2.37 | 10.26 | 97.36 |
| MP | 0.12 | 0.37 | 3 | 0.08 | 0.29 | 2 | 0.09 | 0.31 | 3 |

*Number of observations = 1,979*

***Table 4: Correlation among variables***

| | E | FSS | NE | CP | CE | PP | MP |
|---|---|---|---|---|---|---|---|
| E | 1 | | | | | | |
| FSS | 0.113*** | 1 | | | | | |
| NE | 0.028 | -0.014 | 1 | | | | |
| CP | 0.329*** | -0.003 | -0.013 | 1 | | | |
| CE | -0.007 | -0.052** | -0.007 | -0.202*** | 1 | | |
| PP | 0.254*** | 0.043* | 0.005 | 0.365*** | -0.078*** | 1 | |
| MP | 0.063*** | 0.012 | -0.035 | -0.092*** | 0.345*** | -0.036 | 1 |

*Statistical significance: * p-value <0.10, ** p-value <0.05, *** p-value <0.01*
*Number of observations = 1,979*

***Table 5: Logistic regression results predicting competition outcomes***

|  | ***b*** | OR | Std Err | Z | p>|z| | ***b***Std$_X$ |
|---|---|---|---|---|---|---|
| FSS | 0.013*** | 1.013 | 0.002 | 5.218 | 0.000 | 0.322 |
| NE | 0.468* | 1.597 | 0.272 | 1.716 | 0.086 | 0.093† |
| CP | 0.178*** | 1.195 | 0.015 | 11.648 | 0.000 | 0.655 |
| CE | 0.036** | 1.037 | 0.018 | 1.998 | 0.046 | 0.121 |
| PP | 0.030*** | 1.030 | 0.006 | 4.880 | 0.000 | 0.313 |
| MP | 0.618*** | 1.855 | 0.177 | 3.486 | 0.000 | 0.192 |
| Constant | -2.696*** | - | 0.192 | -14.060 | 0.000 | |

*Dependent variable: competition outcome; method of estimation: logistic regression; $b$ = raw coefficient; OR= Odds Ratio (exp b); z = z-score for test of $b$=0; p>|z| = p-value for z-test; $b$Std$_X$= X standardized coefficient.*

*Statistical significance: * p-value <0.10, ** p-value <0.05, *** p-value <0.01.*

*Number of observations = 1,979; LR chi2(6) = 279.40; Prob > chi2 = 0.0000; Log likelihood = -948.6203; Pseudo R2 = 0.1284; mean VIF=1.11*

*† In this case the standardized coefficient is not considered because the explanatory variable is binary.*

***Table 6: Estimated probabilities of success in competitions on the basis of applicant's FSS and career experience in the same university as the committee president, assuming mean values for other variables***

| | | Years career with president | | | | |
|---|---|---|---|---|---|---|
| | | 0 | 1 | 2 | 5 | 10 |
| FSS | 100 | 0.231 | 0.264 | 0.300 | 0.422 | 0.640 |
| | 50 | 0.135 | 0.158 | 0.183 | 0.276 | 0.481 |
| | 10 | 0.085 | 0.100 | 0.117 | 0.185 | 0.356 |

*Number of observations = 1,979*

***Table 7: Classification of competitions in function of outcome and adequacy of the selection committee***

| Committee/Outcome | | As expected | Not expected |
|---|---|---|---|
| | Frequency | 74 | 77 |
| Adequate | Expected frequency | 67 | 84 |
| | Chi2 contribution | 0.7 | 0.6 |
| | Frequency | 32 | 56 |
| Not adequate | Expected frequency | 39 | 49 |
| | Chi2 contribution | 1.3 | 1.0 |

*Number of observations = 239*

*Table 8: Classification of competitions on the basis of outcome and north-south geographic area*

| Geographic area/Outcome | | As expected | Not expected |
|---|---|---|---|
| North | Frequency | 58 | 68 |
| | Expected frequency | 56.3 | 69.7 |
| | Chi2 contribution | 0.1 | 0.0 |
| South | Frequency | 26 | 36 |
| | Expected frequency | 27.7 | 34.3 |
| | Chi2 contribution | 0.1 | 0.1 |

*Number of observations = 188*

***Table 9: Competition outcomes by UDA.***

| UDA | Competition outcome | | Total competitions |
|---|---|---|---|
| | As expected | Not expected | |
| Mathematics and computer science | 17 (70.8) | 7 (29.2) | 24 |
| Physics | 7 (38.9) | 11 (61.1) | 18 |
| Chemistry | 8 (36.4) | 14 (63.6) | 22 |
| Earth sciences | 4 (100.0) | - | 4 |
| Biology | 9 (45.0) | 11 (55.0) | 20 |
| Medicine | 20 (40.0) | 30 (60.0) | 50 |
| Agricultural and veterinary sciences | 4 (40.0) | 6 (60.0) | 10 |
| Civil engineering and architecture | 5 (45.5) | 6 (54.5) | 11 |
| Industrial and information engineering | 28 (38.4) | 45 (61.6) | 73 |
| Pedagogy and psychology | 1 (25.0) | 3 (75.0) | 4 |
| Economics and statistics | 3 (42.9) | 4 (57.1) | 7 |
| Total | 106 (43.6) | 137 (56.4) | 243 |